\documentclass[seceq,supplement]{ptptex}
\title{From fluctuations in hydrodynamics to nonequilibrium thermodynamics}

\author{%       %Use \scshape for the family name.
Giovanni \textsc{Jona-Lasinio}}

\inst{%     %Affiliation, neglected when [addenda] or [errata].
Dipartimento di Fisica, ``Sapienza'', Universit\`a di Roma, \\
Piazzale A. Moro 2, Roma 00185, Italy\\
Istituto Nazionale di Fisica Nucleare, Sezione di Roma 1, \\
Roma 00185, Italy}

\abst{%         %This abstract is neglected when [addenda] or [errata].
This paper reports on a macroscopic fluctuation theory developed over
the last ten years in collaboration with L. Bertini, A. De Sole,
D. Gabrielli and C. Landim. This theory has been inspired by and tested on
stochastic models of interacting particles (stochastic
lattice gases). It is the basis for a new approach to the study
of stationary non equilibrium states  applicable to a large class 
of systems. This overview emphasizes general ideas and for the details
I refer to the published papers.}

\begin{document}

\maketitle

\section{Introduction}
I will start with a question: why is the theory of irreversible processes
so much more difficult than the theory of equilibrium phenomena? 
Here are some reasons.

\medskip

In equilibrium we do not have to solve any equation of motion and the 
Gibbs distribution provides the basis for the calculation of macroscopic
quantities and their fluctuations. In nonequilibrium we cannot bypass
the dynamics even in the study of stationary states which we may consider
as the simplest beyond equilibrium. It is easy to write
down the equation for the stationary ensemble once we know the microscopic
dynamics, but the solution is in general very hard to find. 
Out of equilibrium therefore the basic object to define analogs of entropy 
or thermodynamic potentials which we need to construct a natural extension
of thermodynamics, is not immediately available. 

\medskip

Since the first attempts to develop a nonequilibrium thermodynamics a guiding 
idea has been that of local equilibrium. This means the following. One assumes
that locally on the macroscopic scale it is possible to define variables
like density, temperature, chemical potentials... which vary smoothly on 
this scale. Microscopically this implies that locally the system reaches
equilibrium in a time which is short compared to the times typical of
macroscopic evolutions as described for example by hydrodynamics. So what
characterizes situations in which this description applies is a separation of 
scales both in space and time. 

\medskip

A theoretical  laboratory where to test this idea is provided by stochastic 
lattice gases which in the last decades have been intensely studied. For several
models local equilibrium has been proved and hydrodynamic evolution
equations have been derived. The microscopic dynamics of these models is 
different from the Hamiltonian dynamics that one would assume for a classical 
real gas but the hope is that the macroscopic behavior of a system,  at least
for certain intervals of time, be qualitatively independent of this
assumption. We expect that general properties like the type and number
of conservations laws should be the most relevant features. 

\medskip

An important achievement in the study of lattice gases has been the
analysis of hydrodynamical fluctuations that is the estimate of the 
probability that the evolution of a macroscopic variable, e.g. the 
density, deviates from a solution of hydrodynamics. This suggests 
a dynamical approach to the estimate of the invariant measure in analogy
with the Freidlin-Ventzell theory for stochastic differential equations
in finite dimension. Let us give for convenience of the reader a
sketch of the Freidlin-Ventzell scheme. \cite{FV} 

\medskip

Consider a stochastic differential equation 
\begin{eqnarray}
%\nonumber
dx_t=b(x_t)dt + \epsilon \sigma dw_t,
\end{eqnarray}
where the vector field $b$ is the drift and $\sigma$ the diffusion matrix.
One is  interested in the limit $\epsilon \rightarrow 0$.
Then the following holds: the probability that the solution stay
close to a trajectory $\phi_t$ in a fixed time interval $[0,T]$ is
\begin{eqnarray}
%\nonumber
P(x_t \simeq \phi_t) \simeq \exp {(- \frac 1{\epsilon^2} I_T (\phi_t))},
\end{eqnarray}
where
\begin{eqnarray}
\label{1.3}
I_T (\phi_t)=\frac 12 \int_0^T dt (\dot \phi - b(\phi_t)){\sigma^{-2}}
(\dot \phi - b(\phi_t)).
\end{eqnarray}
$I_T$ is called the large deviation functional.

\medskip      

From this one obtains the estimate of the stationary distribution in the
neighborhood of an equilibrium point
\begin{eqnarray}
%\nonumber
P(x) \simeq \exp {(- \frac 1{\epsilon^2} V(x))},
\end{eqnarray} 
where $V(x)=I_\infty (\phi^*_t)$ with $\phi^*_t$ a trajectory connecting
the equilibrium point to $x$ and minimizing $I_\infty$. If there are
several equilibrium points or attractors the theory can be easily
extended.

\medskip

{\sl {The lesson we learn  is that in the limit of small noise
an estimate of the stationary distribution is reduced to the solution
of a variational problem.}} 

\medskip

Equation (1.4) reminds of the Einstein theory of equilibrium thermodynamic
fluctuations \cite{E,LL} which states that the probability of a 
fluctuation from equilibrium in a macroscopic region of volume $V$ is 
proportional to 
$$ 
\exp\{V\Delta S / k\},
$$ 
where $\Delta S$ is the variation of entropy density calculated along
a reversible transformation creating the fluctuation and $k$ is the
Boltzmann constant. This theory is well established and has received a
rigorous mathematical formulation in classical equilibrium statistical
mechanics. \cite{La} The analogy with Einstein formula  is especially
suggestive: establishing estimates of this kind in nonequilibrium leads 
naturally to the identification of the exponent with a nonequilibrium 
thermodynamic function. Clearly the role of the small parameter $\epsilon$ 
is taken by the inverse of the volume. This is the starting point of the
theory reviewed in this paper.
For related approaches see. \cite{DRe,MNS} 

\section{Basic equations and the large deviation functional}
The following equations characterize the macroscopic behaviour of a wide class
of stochastic lattice gases and agree with the phenomenological equations 
used so far in nonequilibrium thermodynamics near equilibrium, see e.g. 
the book.\cite{FI}

\medskip

\begin{itemize}
\item[1.]
\emph{The macroscopic state is completely described by
the local density $\rho=\rho(t,x)$ and the associated current $j=j(t,x)$.}

\item[2.]\emph{The macroscopic evolution is given by the continuity equation
\begin{equation}
\label{2.1}
\partial_t \rho + \nabla\cdot j = 0,
\end{equation}
together with the constitutive equation
\begin{equation}
\label{2.2}
j = J(\rho) = - D(\rho) \nabla\rho + \chi(\rho) E,
\end{equation}
where the \emph{diffusion coefficient} $D(\rho)$ and the \emph{mobility}
$\chi(\rho)$ are $d\times d$ positive matrices.
The transport coefficients $D$ and $\chi$ satisfy the local Einstein
relation
\begin{equation}
\label{ein_rel}
D(\rho) = \chi(\rho) \, f_0''(\rho),
\end{equation}
where $f_0$ is the equilibrium free energy of the homogeneous system.}
\end{itemize}

\medskip

The equations \eqref{2.1}--\eqref{2.2} have to be supplemented by the
appropriate boundary conditions on $\partial\Lambda$ due to the
interaction with the external reservoirs. If
$\lambda_0(x)$, $x\in\partial \Lambda$, is the chemical potential of
the external reservoirs, these boundary conditions are
\begin{equation}
\label{2.3}
f_0'\big(\rho(x) \big) = \lambda_0(x), \qquad\qquad x\in\partial
\Lambda.
\end{equation}

\medskip

Some comments are in order.
There are no restrictions on the values of the chemical potential or
on the nonlinearity 
so that situations far from equilibrium or from a stationary state 
are included. 
We denote by $\bar\rho=\bar\rho(x)$, $x\in\Lambda$, the stationary
solution, assumed to be unique, of equations \eqref{2.1} to (2.4).

In the constitutive equation 
\eqref{2.2} the external field appears linearly which microscopically
means a field of order $ 1/N$ acting on each particle, where $N$ is
the linear dimension of the lattice. A field of order $1$ would destroy
the diffusive character of the evolution giving rise to a
hyperbolic evolution equation of first order. 

The transport coefficients
$D$ and $\chi$ depend in general on the density $\rho$.
In the case the system has more than one component, say $n$, the
diffusion coefficient $D$ and the mobility become $nd\times nd$
matrices. Moreover, in view of Onsager reciprocity, the matrix $\chi$
is symmetric both in the space and in the component indices while $D$
is symmetric only in the space indices. In such a case the local Einstein
relation \eqref{ein_rel} is $D =  \chi \, R$ where
$R_{ij} = \partial_{\rho_i} \partial_{\rho_j} f_0$
does not depend on the space indices.

Summarizing, in the context of stochastic lattice gases, \eqref{2.1} and
\eqref{2.2} describe the evolution of the density in the
diffusive scaling limit, see e.g.\ \cite{KL,S}.
The validity of the local Einstein relationship \eqref{ein_rel} can be
deduced from the local detailed balance of the underlying microscopic
dynamics, see e.g.\ \cite{S} 

\medskip

So far the analysis of the stochastic models can be viewed as a
confirmation of the phenomenological schemes used by physicists and
chemists for a long time. The really new information comes from
the study of large deviations from the above deterministic macroscopic 
description.

\medskip

For a wide class of models the following holds.
The stationary measure $P_{st}$ admits a principle of large deviations
describing the fluctuations of the thermodynamic variable appearing
in the hydrodynamic equation. This means the following. 
The probability that in a lattice of  $N^d$ points   the evolution of 
the so-called empirical density $\rho_N (X_t) = \frac{1}{N^d} \sum_{x\in\Lambda_N} 
\eta_x(N^2 t) \: \delta \left(u - \frac xN \right)$, where $\eta_x$ is 
the number of particles at site $x$ and $X_t$ is the microscopic configuration,
deviates from the solution of the hydrodynamic equation and is close
to some trajectory ${\hat{\rho}}(t)$, 
is exponentially small and of the form
\begin{equation}
\label{LD}
P_{st}\left( 
\rho_N(X_t) \sim\hat{\rho}(t), t\in [t_1, t_2]\right) 
\approx e^{-N^d[{\mathcal F}(\hat \rho({t_1})) + I_{[t_1,t_2]}(\hat{\rho})] }, 
\end{equation}
where $I(\hat{\rho})$ is a functional which vanishes if
${\hat{\rho}}(t)$ is a solution of \eqref{2.1}--\eqref{2.2} and ${\mathcal F}
(\hat \rho({t_1}))$
is the  cost to produce the initial density profile
${\hat{\rho}}({t_1})$.   We normalize it so that ${\mathcal F}(\bar\rho)=0$.
The functional $I(\hat{\rho})$ represents the extra cost necessary to
follow the trajectory ${\hat{\rho}}(t)$. Finally $\rho_N(X_t) \sim
\hat{\rho}(t)$ means closeness in some metric and $\approx$ denotes
logarithmic equivalence as $N\to\infty$.  

\medskip

A simple argument based on time reversal \cite{BDGJL2} now shows that 
\begin{eqnarray} 
{\mathcal F}(\rho)= \inf_{\hat \rho}  I_{[-\infty,0]}(\hat \rho),  
\end{eqnarray}
where the $\inf$ has to be taken over all trajectories connecting the stationary
state $\bar \rho$ to the profile $\rho$. 
The same argument identifies the minimizing trajectory, see later. 
The quantity ${\mathcal F}(\rho)$
is an infinite dimensional analog of the $V(x)$ in Freidlin-Ventzell
theory and has to be identified with a non equilibrium thermodynamic function.  
The fluctuations we are considering take place at constant 
temperature and volume so that it is reasonable to identify ${\mathcal F}(\rho)$
with the variation of the nonequilibrium free energy to produce the profile
$\rho$. If the stationary state is an equilibrium state the Einstein theory 
is recovered as it will be discussed later.

\medskip

The explicit formula of the functional $I_{[t_1,t_2]}(\hat{\rho})]$ 
is
\begin{equation}
\label{I=}
I_{{[T_1,T_2]}} (\hat\rho) =
\frac 14 \int_{{T_1}}^{{T_2}}\!dt
\: \Big\langle
\big[ \partial_t \hat\rho  + \nabla \cdot J(\hat\rho)
\big] \, K(\hat\rho)^{-1}
\big[ \partial_t \hat\rho  + \nabla \cdot J(\hat\rho) \big] \Big\rangle,
\end{equation}
where the positive operator $K(\hat\rho)$, the analog of $\sigma^2$ in
Freidlin-Ventzell theory,  is defined on functions $u$ 
vanishing at the boundary $\partial \Lambda$ by
$K(\hat\rho) u = - \nabla \cdot\big( \chi(\hat\rho) \nabla u \big)$.
The expression (2.7) is similar to (1.3), i.e. the fluctuations
of the thermodynamic variables of a stochastic lattice gas are
formally those of a stochastic partial differential equation where
the deterministic part is given by hydrodynamics. These equations
are known as {\sl fluctuating hydrodynamics} and are often used in 
phenomenological  
calculations. One has to be careful however because it is often difficult to
attribute a precise mathematical meaning to stochastic nonlinear partial 
differential equations. Divergences akin to those in quantum field theory
appear in dimension greater than $1$.  

\medskip

For a heuristic derivation of (2.7) see  \cite{BDGJL2}. There is
factor $1/2$ of difference due to a slightly different definition
of the transport coefficients $D$ and $\chi$.
\section{The Hamilton-Jacobi equation and the interpretation of 
$\mathcal F$}
As shown in \cite{BDGJL1,BDGJL2} the functional $\mathcal F$, as defined
in (2.6), is the maximal
solution of the infinite dimensional Hamilton-Jacobi equation
\begin{equation}
\label{HJeq}
\Big\langle  \nabla \frac{\delta\mathcal F}{\delta\rho} \cdot \chi(\rho)
\nabla \frac{\delta\mathcal F}{\delta\rho} \Big\rangle -
\Big\langle  \frac{\delta\mathcal F}{\delta\rho}
\: \nabla \cdot J(\rho) \Big\rangle  = 0,
\end{equation}
where, for $\rho$ that satisfies \eqref{2.3}, $\delta\mathcal F /
\delta\rho$ vanishes at the boundary of $\Lambda$.  At the macroscopic
level this condition reflects the fact that we consider variations of
the density that do not change the boundary values. 
The arbitrary additive constant on the maximal solution of
\eqref{HJeq} is determined by the condition $\mathcal F(\bar\rho)=0$. By
maximal solution we mean that any solution $F(\rho)$ to \eqref{HJeq} (satisfying
$F(\bar\rho)=0$) is a lower bound for $\mathcal F$. 
In nonequilibrium we expect generically  $\mathcal F$ to 
depend nonlocally on the density.

\medskip

It is easy to see that the Hamilton-Jacobi equation implies that
$\nabla J$ can be decomposed in such a way that the hydrodynamic equations
can be written

\begin{eqnarray}
\partial_t \rho &=&-\nabla J(\rho)=   
\nabla \cdot 
\Big( \chi(\rho) \nabla \frac {\delta \mathcal F}{\delta \rho} \Big) 
+ \mathcal{A}(\rho),
\end{eqnarray}
where $\mathcal{A}(\rho)$ satisfies the orthogonality condition 
\begin{eqnarray}
\Big\langle  \frac {\delta \mathcal F}{\delta \rho} \,,\, \mathcal{A}(\rho)
\Big\rangle = 0.
\end{eqnarray}

This decomposition confirms the interpretation of $\mathcal F$ as a 
nonequilibrium free energy. In fact the first term is the dissipative
part of the equation which determines the relaxation to the 
stationary state due to the thermodynamic force 
$\nabla \frac {\delta \mathcal F}{\delta \rho}$ while $\mathcal{A}(\rho)$
does not contribute to the spontaneous relaxation. 
$\mathcal{A}(\rho)$ 
is minus the divergence of the non-dissipative part of the current and
vanishes in the stationary state. 
It can be shown that
\cite{BDGJL2} the hydrodynamics associated to the time reversed microscopic
process can be written
\begin{eqnarray}
\partial_t \rho &=&-\nabla J^*(\rho)=  
\nabla \cdot 
\Big( \chi(\rho) \nabla \frac {\delta \mathcal F}{\delta \rho} \Big) 
- \mathcal{A}(\rho),
\end{eqnarray}
where $J^*(\rho)$ is the current associated to the time reversed process.
By summing equations (3.2) and (3.4) we obtain the nonequilibrium
fluctuation-dissipation relation
\begin{equation}
\label{2.7}
\nabla(J^*(\rho)+J(\rho)) = - 2\nabla\chi(\rho) \nabla \frac{\delta\mathcal F}{\delta\rho}.  
\end{equation}
Actually a similar relation holds for the currents \cite{BDGJL6}
\begin{equation}
\label{2.7}
J^*(\rho)+J(\rho) = - 2\chi(\rho) \nabla \frac{\delta\mathcal F}{\delta\rho}.  
\end{equation}

%\medskip

We can now identify the optimal (minimizing) trajectory in (2.6).
Let $\mathcal F$ be the maximal solution of the Hamilton-Jacobi equation and
$J^*$ as in \eqref{2.7}.
Fix a time interval ${[T_1,T_2]}$ and a
path $\hat \rho (t)$, $t\in {[T_1,T_2]}$. We claim that
\begin{eqnarray}
\label{I=I*}
\nonumber
&& I_{{[T_1,T_2]}} (\hat\rho) = \mathcal F\big(\hat\rho({T_2})\big) 
-\mathcal F\big(\hat\rho({T_1}) \big)
\\
&& \qquad\qquad
+ \frac 14 \int_{{T_1}}^{{T_2}}\!dt\:
\: \Big\langle
\big[ \partial_t \hat\rho  - \nabla \cdot J^*(\hat\rho)
\big] \, K(\hat\rho)^{-1}
\big[ \partial_t \hat\rho  - \nabla \cdot J^*(\hat\rho) \big]
\Big\rangle, \qquad
\end{eqnarray}
as can be shown by a direct computation using (2.7),
the Hamilton-Jacobi equation \eqref{HJeq} and (3.6). 
From the identity \eqref{I=I*} we immediately deduce
that the optimal path for the variational problem 
(2.6) is the time reversal of the solution to (3.4) with initial condition
$\rho$.

\medskip

We next show that according to definition (2.6), the free energy
$\mathcal F(\rho)$ is equal to the work done by the thermodynamic force
on the system along the optimal trajectory $\rho^*$.  
Indeed, by the above identification of $\rho^*$ 
\begin{eqnarray*}
\mathcal F(\rho)-\mathcal F(\bar\rho) 
&= &\int_{-\infty}^0 \!dt \, \Big\langle \frac{\delta\mathcal F}{\delta \rho},
\partial_t\rho^* \Big\rangle
\\
&=& \int_{-\infty}^0 \!dt\, \Big\langle \frac{\delta\mathcal F}{\delta \rho},
\nabla \cdot J^*(\rho^*) \Big\rangle
=\int_{-\infty}^0 \!dt \, \Big\langle (-J^*(\rho^*))\cdot \nabla \frac{\delta\mathcal F}{\delta \rho}\Big\rangle.
\end{eqnarray*}

\medskip
 
The  decompositions (3.2)-(3.4) remind of the electrical
conduction in presence of a magnetic field \cite{BDGJL3}.
Consider the motion of electrons in a conductor: a simple 
model is given by the effective equation \cite{AM}, 
\begin{equation}
{\dot {\bf p}} = -e \Big( {\bf E} 
+ \frac{1}{mc} {\bf p} \wedge {\bf H} \Big)
- \frac{1}{\tau} {\bf p},
\label{ec}
\end{equation}
where ${\bf p}$ is the momentum, $e$ the electron charge, 
${\bf E}$ the electric field, ${\bf H}$ the magnetic field, 
$m$ the mass, $c$ the velocity of the light, and $\tau$ the relaxation
time. The dissipative term ${\bf p}/\tau$ is orthogonal to the
Lorenz force ${\bf p} \wedge {\bf H}$. 
We define time reversal as the transformation ${\bf p}\mapsto -{\bf p}$, 
${\bf H}\mapsto -{\bf H}$. The time reversed evolution is given by
\begin{equation}
{\dot {\bf p}} = e \Big( {\bf E} 
+\frac{1}{mc} {\bf p} \wedge {\bf H} \Big)
- \frac{1}{\tau} {\bf p}.
\label{ec*}
\end{equation}
Let us consider in particular the Hall effect where we have conduction
along a rectangular plate immersed in a perpendicular magnetic field
$H$ with a potential difference across the long side.  The magnetic
field determines a potential difference across the short side of the
plate.  In our setting on the contrary it is the difference in
chemical potentials at the boundaries that introduces in the equations
a non-dissipative term.

%\medskip

\section{Equilibrium states and macroscopic reversibility}
In this section we consider the notion of \emph{equilibrium}
from the standpoint of \emph{nonequilibrium} \cite{BDGJL7}.
We define the system to be in
\emph{equilibrium} if and only if the current in the stationary profile
$\bar\rho$ vanishes, i.e.\ $J(\bar\rho) = 0$.
Nonetheless, in
presence of external (electric) fields and boundary reservoirs an
equilibrium state can be highly inhomogeneous.  An example of such a
situation is provided by sedimentation equilibrium in gravitational
and centrifugal fields.  In spite of this, the free energy is a local
function of the state variables and coincides locally with the
equilibrium free energy in absence of external fields and boundary
driving. In particular there are no macroscopic correlations.

\medskip

For an equilibrium state, characterized
by a constant or non constant stationary profile $\bar\rho(x)$ such that
$J(\bar\rho)=0$, the free energy functional $\mathcal F$ is obtained as follows.
Let
\begin{equation*}
f(\rho,x) =
\int_{\bar\rho(x)}^\rho \!dr \int_{\bar\rho(x)}^r \!dr' \: f_0''(r')
= f_0(\rho) - f_0(\bar\rho(x)) - f_0'\big(\bar\rho(x)\big) \big[ \rho
-\bar\rho(x)\big], 
\end{equation*}
where, we recall, $f_0 (\rho)$ is the equilibrium free energy density.
It is easy to show \cite{BDGJL7} that 
the maximal solution of the Hamilton-Jacobi equation is 
\begin{equation}
\label{Floc}
\mathcal F(\rho) = \int_\Lambda \!dx \: f\big( \rho(x), x\big).
\end{equation}
We emphasize that 
the above argument depends crucially on the 
structure \eqref{2.2} of the current and on validity
of the local Einstein relation \eqref{ein_rel}.

\medskip

The condition $J(\bar\rho)=0$ can be shown \cite{BDGJL7} to be
equivalent to $J^*(\rho) = J(\rho)$ for any profile $\rho$. 
We call the latter condition {\sl macroscopic reversibility}.
The notion of macroscopic reversibility does not
imply that an underlying microscopic model satisfies the detailed
balance condition. 
Indeed, as it has been shown by explicit examples
\cite{GJL1,GJL2}, there are non reversible microscopic models which are
macroscopically reversible. If the system is macroscopically reversible
the optimal trajectory to create a fluctuation is the time reversal
of  the relaxation trajectory solution of \eqref{2.1}-\eqref{2.2}.
In \cite{BDGJL2} we called this symmetry the Onsager-Machlup time
reversal symmetry \cite{ONS1, OMA}.

We have defined the macroscopic reversibility as the identity between
the currents $J(\rho)$ and $J^*(\rho)$. We emphasize that this is not
equivalent to the identity between $\nabla \cdot J(\rho)$ and $\nabla
\cdot J^*(\rho)$.  Indeed, it is possible to give  examples \cite{BDGJL7} of
a non reversible system, i.e.\ {with} $J(\bar\rho) \neq 0$, such that 
the optimal
trajectory for the variational problem (2.6) is the time
reversal of the solution to the hydrodynamic equation
\eqref{2.1}-\eqref{2.2}.

\medskip

Our analysis of equilibrium so far does not include magnetic fields but can
be extended to cover this situation. It can be done by distinguishing
dissipative and non dissipative currents: a natural definition
of equilibrium is then the vanishing of  the dissipative currents.  

\section{Long range correlations}
Space correlations extending over macroscopic distances appear to be
a generic feature of stationary nonequilibrium states, a fact  
known since a long time \cite{S1,DKS}. 
The ultimate reason for these correlations seems to be connected
with the violation of time reversal as they can appear
also in equilibrium states if the microscopic dynamics {\sl strongly}
violates time reversal invariance \cite{BJL, BDGJL9}. An interesting question
is to what extent long range correlations make nonequilibrium akin
to critical phenomena.

In our approach, since we are not limited to the vicinity of the stationary
state we can obtain the (nonequilibrium) density
correlations functions of arbitrary order in terms of the functional
derivatives of $\mathcal F$. In general the functional $\mathcal F$ cannot be
written in a closed form, but - by a suitable perturbation theory on the
Hamilton-Jacobi equation \eqref{HJeq} - we can derive such
correlations functions.
In this section we  discuss the two-point correlation and establish a
criterion to decide whether the density fluctuations are positively or
negatively correlated.  Recursive equations for the
correlation functions of any order are established in \cite{BDGJL7}. 
We emphasize that we are concerned only with \emph{macroscopic
  correlations} which are a generic feature of nonequilibrium models.
Microscopic correlations which decay exponentially or as a summable power law 
disappear at the macroscopic level.

\medskip

We introduce the \emph{pressure} functional as the Legendre
transform of free energy $\mathcal F$
\begin{equation*}
\mathcal G(h) = \sup_\rho\big\{\langle h \rho\rangle -\mathcal F(\rho)\big\}.
\end{equation*}
By Legendre duality
{
we have the change of variable formula
$h=\frac{\delta\mathcal F}{\delta\rho},\,\rho=\frac{\delta\mathcal G}
{\delta h}$,
so that
}
the Hamilton-Jacobi equation \eqref{HJeq}
can be rewritten in terms of $\mathcal G$ as
\begin{equation}
\label{s2}
\Big\langle \nabla h \cdot
\chi\Big(\frac{\delta\mathcal G}{\delta h} \Big)\nabla h
\Big\rangle
- \Big\langle \nabla h \cdot
 D \Big(\frac{\delta\mathcal G}{\delta h}\Big) \nabla\frac{\delta\mathcal
  G}{\delta h}  {-} \chi\Big(\frac{\delta\mathcal G}{\delta h}\Big)E
\Big\rangle = 0,
\end{equation}
where $h$ vanishes at the boundary of $\Lambda$.
As for equilibrium systems, $\mathcal G$ is the generating functional of the correlation
functions. In particular, by defining
\begin{equation*}
C(x,y) = \frac{\delta^2\mathcal G(h)}{\delta h(x) \delta h(y)} \,{\Big|_{h=0}}.
\end{equation*}
we have, since $\mathcal F$ has a minimum at $\bar\rho$,
\begin{equation*}
\mathcal G(h) = \langle h,\bar\rho\rangle + \frac12 \langle
h,Ch\rangle + o(h^2),
\end{equation*}
or equivalently
\begin{equation*}
\mathcal F(\rho) = \frac12
\langle(\rho-\bar\rho),C^{-1}(\rho-\bar\rho)\rangle +
o((\rho-\bar\rho)^2).
\end{equation*}

By expanding the Hamilton-Jacobi equation \eqref{s2}
to the second order in $h$, and using that
$\delta \mathcal G /\delta h(x) = \bar\rho (x) + C h(x) +o(h^2)$,
we get the following equation for $C$
\begin{equation}
\label{s4}
\Big\langle \nabla h \cdot \Big[
\chi(\bar\rho)\nabla h -
\nabla(D(\bar\rho)Ch)+\chi^\prime(\bar\rho)(Ch)E \Big]\Big\rangle = 0.
\end{equation}

We now make the change of variable
$$
C(x,y)=C_{\mathrm{eq}}(x)\delta(x-y)+B(x,y),
$$
where $C_{\mathrm{eq}}(x)$ is the equilibrium covariance. By using
\eqref{ein_rel} we deduce that
$$
C_{\mathrm{eq}}(x)= D^{-1}(\bar\rho(x))\chi(\bar\rho(x)).
$$
Equation \eqref{s4} for the correlation function then gives the
following equation for $B$
\begin{equation}\label{s5}
\mathcal L^\dagger B(x,y) = \alpha(x)\delta(x-y),
\end{equation}
where $\mathcal L^\dagger$ is the formal adjoint of the
elliptic operator $\mathcal L=L_x+L_y$ given by, using the usual convention
that repeated indices are summed,
\begin{equation}\label{lx}
L_x = D_{ij}(\bar\rho(x))\partial_{x_i}\partial_{x_j} +
\chi^\prime_{ij}(\bar\rho(x)) E_j(x)\partial_{x_i},
\end{equation}
and
\begin{equation*}
\alpha(x) = \partial_{x_i}\big[
\chi^\prime_{ij}\big(\bar\rho(x)\big)\,
D^{-1}_{jk}\big(\bar\rho(x) \big)\bar J_{k}(x) \big],
\end{equation*}
where we recall
$\bar J= J (\bar\rho)=- D(\bar\rho(x))\nabla\bar\rho(x)
+\chi(\bar\rho(x))E(x)$
is the macroscopic current in the stationary profile.

In equilibrium systems $\bar J=0$ so that we have
$\alpha=0$, hence $B=0$, namely there are no long range correlations
and $C(x,y)=C_{\mathrm{eq}}(x)\delta(x-y)$.
Moreover, since $\mathcal L$ is an elliptic operator 
(i.e. it has a
negative kernel), the sign of $B$ is determined by the sign of $\alpha$:
if $\alpha(x) \ge 0,\,\forall x$, then $B(x,y) \le 0,\, \forall x,y$,
while if $\alpha(x) \le 0,\,\forall x$, then $B(x,y) \ge 0,\, \forall
x,y$. 
For example, consider the following special case. The system is one-dimensional,
$d=1$, the diffusion coefficient is constant, i.e.\ $D(\rho)=D_0$,
the mobility $\chi(\rho)$ is a quadratic function of $\rho$, and there
is no external field, $E=0$. Then
\begin{equation}
\label{opp}
B(x,y) = - \frac 1{2D_0} \chi'' (\nabla \bar\rho)^2 \Delta^{-1}(x,y),
\end{equation}
where $\Delta^{-1}(x,y)$ is the Green function of the Dirichlet
Laplacian. Two well studied models, the symmetric exclusion process,
where $\chi(\rho)=\rho(1-\rho)$, and the KMP process \cite{BGL}, where
$\chi(\rho)=\rho^2$, meet the above conditions. Then \eqref{opp} shows that 
their correlations have opposite signs. 

\section{Thermodynamics of currents}
In nonequilibrium a very important observable is the current flux. This quantity
gives information that cannot be recovered from the density because
from a density trajectory we can determine the current trajectory only
up to a divergence free vector field. 
To discuss the current fluctuations, we introduce a vector-valued
observable $\mathcal J_N(\{X_\sigma$, $0\le \sigma\le \tau\})$ of the
microscopic trajectory $X_t$ which measures the local net flow of particles
and satisfies formally the microscopic continuity equation
\begin{equation}
\nonumber 
\partial_t \rho_N + \nabla_N \cdot \mathcal J_N = 0,
\end{equation}
where $\nabla_N$ is the
gradient on the lattice. For the details see \cite{BDGJL5,BDGJL6}.  
As in the case of the density, for stochastic lattice gases, 
we are able to derive
a dynamical large deviations principle for the current. If
$P_{X_0}$ stands for the probability $P_{st}$ conditioned on the
initial microscopic state $X_0$,
given a vector field $j:[0,T]\times \Lambda \to \mathbb R^d$, we have
\begin{equation}
\label{f1a}
P_{X_0} \big( \mathcal J_N (X) \approx j (t,u) \big) 
\sim \exp\big\{ - N^d \, \mathcal I_{[0,T]}(j)\big\},
\end{equation}
where the rate functional is 
\begin{equation}
\label{Ica}
\mathcal I_{[0,T]}(j)\;=\; \frac 14 \int_0^T \!dt \,
\big\langle [ j - J(\rho)  ], \chi(\rho)^{-1}
[ j - J(\rho) ] \big\rangle.
\end{equation}
We recall that 
\begin{equation*}
J(\rho) = -  D(\rho) \nabla \rho  + \chi(\rho) E\; .
\end{equation*}
In (6.2) $\rho=\rho(t,u)$ is the solution of the continuity
equation $\partial_t\rho +\nabla\cdot j =0$ with the initial condition
$\rho(0)=\rho_0$ associated to $X_0$. The rate functional vanishes if
$j=J(\rho)$.

\medskip

Among the many problems we can discuss within
this theory, the fluctuations of the time average of the
current $\mathcal J_N$ over a large time interval have been analysed.  
This question was
addressed by Bodineau and Derrida in \cite{BD} in one space dimension 
by postulating an {\sl ``additivity principle''} which relates the fluctuation 
of the time averaged current in the whole system to the fluctuations in
subsystems. However their approach does not always apply. In fact  
the probability of observing a given
divergence free time averaged fluctuation $J$ can be described by a
functional $\Phi(J)$ which we characterize, in any dimension, in terms
of a variational problem for the functional $\mathcal I_{[0,T]}$
\begin{equation}
\label{limTa}
\Phi (J)
 = \lim_{T\to\infty} \; \inf_{j} 
\frac 1T \; \mathcal I_{[0,T]} (j)\;,
\end{equation}
where the infimum is carried over all paths $j=j(t,u)$ having time
average $J$. The static additivity principle postulated in \cite{BD}
gives the correct answer only under additional hypotheses which are
not always satisfied. 
Let us denote by $U$ the functional obtained by restricting the
infimum in \eqref{limTa} to divergence free current paths $j$, i.e.\ 
\begin{equation}
\label{Ua}
U(J) =  \inf_{\rho} \frac 14 
\big\langle [ J - J(\rho)  ], \chi(\rho)^{-1}
[ J - J(\rho) ] \big\rangle,
\end{equation} 
where the infimum is carried out over all the density profiles
$\rho=\rho(u)$ satisfying the appropriate boundary conditions. From
\eqref{limTa} and \eqref{Ua} it follows that $\Phi \le U$. 
In one space dimension the functional $U$ is the one introduced in
\cite{BD}. While $\Phi$ is always convex the functional $U$ 
may be non convex. In such a case $U(J)$ underestimates the
probability of the fluctuation $J$.  In \cite{BDGJL5, BDGJL6} we interpreted 
the lack of convexity of $U$, and more generally the strict inequality
$\Phi<U$,  as a dynamical phase transition. 

There are cases in which $\Phi= U$. Sufficient conditions on 
the transport
coefficients $D$, $\chi$ for the coincidence of $\Phi$ and $U$
can be given \cite{BDGJL6}.
Consider the case when the matrices $D(\rho)$ and
$\chi(\rho)$ are multiples of the identity, i.e., there are strictly
positive scalar functions still denoted by $D(\rho)$, $\chi(\rho)$, so
that $D(\rho)_{i,j}= D(\rho) \delta_{i,j}$, $\chi(\rho)_{i,j}=
\chi(\rho) \delta_{i,j}$, $i,j=1,\dots ,d$. 
Let us first consider the case with no external field, i.e.\ $E=0$; if 
\begin{equation}
\label{c<}
D(\rho) \chi''(\rho) \le D'(\rho) \chi'(\rho), 
\quad \textrm{ for any } \rho, 
\end{equation}
where $'$ denotes the derivative,
then $\Phi=U$. In this case $U$ is necessarily convex. 
Moreover if  
\begin{equation}
\label{c=}
D(\rho) \chi''(\rho) = D'(\rho) \chi'(\rho), 
\quad \textrm{ for any } \rho, 
\end{equation}
then we have $\Phi=U$ for any external field $E$.

To exemplify  situations in which $\Phi < U$ consider the fluctuations of 
the time averaged current for periodic boundary conditions. Two models 
have been discussed so far. The Kipnis--Marchioro-Presutti (KMP) 
model \cite{KMP},
which is defined by a harmonic chain with random exchange of energy
between neighboring oscillators, and the exclusion process.
In the case of the KMP model we have $U(J)= (1/4) J^2/ \chi(m)=
(1/4) J^2/m^2 $, where $m$ is the (conserved) total energy. For $J$ large 
enough, $\Phi(J) < U(J)$.  This
inequality is obtained by constructing a suitable travelling wave
current path whose cost is less than $U(J)$ \cite{BDGJL6}. For $J$ not too large
the additivity principle holds as it has been verified numerically
in \cite{HG}. A similar result has been obtained 
by Bodineau and Derrida \cite{BD1} for the
periodic simple exclusion process with external field.  
For the KMP process this phenomenon is rather
striking as it occurs even in equilibrium, i.e.\ without external field.

The behavior of $\mathcal I$ and $\Phi$ under time reversal
shows that $\Phi$ satisfies a fluctuation relationship akin to 
the Gallavotti-Cohen
theorem for the entropy production \cite{gc,k,ls}.  
The anti-symmetric part of $\Phi$ is equal to the
power produced by the external field and the reservoirs independently
of the details of the model 
\begin{equation}
\label{GC2}
\Phi(J)- \Phi(-J)=\Phi(J)- \Phi^a(J)=
- 2 \langle J ,E \rangle 
+\int_{\partial\Lambda} \!d\Sigma \: \lambda_0 \, J \cdot \hat{n}, 
\end{equation}
the right hand side of this equation
is the power produced by the external field and the
boundary reservoirs (recall $E$ is the external field and $\lambda_0$
the chemical potential of the boundary reservoirs).
From this relationship one derives  a
macroscopic version of the fluctuation theorem for the entropy
production.

\medskip

For recent interesting results obtained from the macroscopic fluctuation
theory in the study of current fluctuations see \cite{ARDLW, DG}.

\section{Conclusions and comparison with other approaches }
The theory developed so far, as emphasized in \cite{BDGJL7}, can be
viewed as a selfcontained macroscopic description of diffusive 
systems out of equilibrium. It allows a clear identification of the
dissipative part and of thermodynamic forces in relaxation
phenomena. Long range space correlations among thermodynamic
variables are a generic consequence of the theory and equations
for the correlation functions of any order have been established
\cite{BDGJL7}. 
Also different regimes in current
fluctuations are predicted that have been characterized as
dynamical phase transitions. These transitions are shown to exist
in the simplest models considered and it is a challenge to discover
them in real systems. 

\medskip

The thermodynamic functionals $\mathcal F$
and $\Phi$ can be calculated from dynamical measurable quantities like
the transport coefficients $D$ and $\chi$. This is a deep difference
with respect to equilibrium thermodynamics where static properties
like specific heats are involved. From the standpoint of nonequilibrium
it is natural to consider also the equilibrium free energy as determined
by the transport coefficients through the Einstein relation.

Another substantial difference
is that in the definition of nonequilibrium thermodynamic functionals
optimal trajectories are involved which are different from
the infinitely slow reversible transformations of classical thermodynamics: 
the optimal trajectories go through
nonequilibrium nonstationary states. In the theory developed so far the boundary
conditions are kept fixed: the study under boundary conditions (chemical
potentials, volume...) which slowly vary on the macroscopic time scale
is a next natural step. 

\medskip

A different very general  approach to the theory of nonequilibrium 
stationary
states was initiated by Oono and Paniconi \cite{OP} and pursued
in the work of Hatano, Hayashi, Sasa, Tasaki.
\cite{HaS,HSa,ST} While in \cite{HaS,HSa} microscopic models underlie
the analysis, in \cite{ST} a guiding idea is to keep as much as possible
the phenomenological character of classical thermodynamics without
reference to an underlying microscopic dynamics. In this work the authors
discuss the operational definition of nonequilibrium thermodynamic observables
in concrete situations and generalize basic operations like decomposition,
combination and scaling of equilibrium thermodynamics to nonequilibrium
states. The possibility of experimental tests is then discussed. 

In more recent papers by Komatsu, Nakagawa 
\cite{KN} and by Komatsu, Nakagawa, Sasa, Tasaki, 
\cite{KNST,KNST1}
the problem of constructing  microscopic ensembles describing stationary states
of both stochastic and Hamiltonian systems, is considered. 
Expressions for the nonequilibrium distribution function are proposed  
either exact or valid up to a certain order
in the parameters keeping the system out of equilibrium.
The key quantity appearing in these expressions is the entropy production.
In \cite{KNST} heat conduction is considered in particular and an extension
of the Clausius and Gibbs relations is derived.
    
A direct comparison between these works and ours 
is not immediately available as we are asking different 
questions but an effort should be made to construct a bridge between 
them.

\medskip

In a recent paper \"Ottinger \cite{OT} compared his own approach \cite{OT1} to
nonequilibrium, called GENERIC (general equation for the nonequilibrium 
reversible-irreversible coupling), with our macroscopic fluctuation
theory discussing aspects where a  correspondence could  
be established. His starting point is a separation in the macroscopic
evolution equations of dissipative and conservative terms which
reminds of our decomposition (3.2). An important difference is 
related to the fact that our free energy out of equilibrium is
generically nonlocal in space and this is connected with the existence of
long range correlations. There is more to understand and
the comparison should be developed further.

\section*{Acknowledgements}

I am very grateful to the organizing committee of the 2009 YKIS Workshop
on nonequilibrium statistical mechanics for the invitation and for the warm
hospitality.

%%%%%%%%%%%%%%%%%%%%%%%%%%%%%%%%%%%%%%%%%%%%%%%%%%%%%%%%%%%%%
% Some macros are available for the bibliography:
%  o for general use
%    \JL : general journals                 \andvol : Vol (Year) Page
%  o for individual journal 
%    \AJ   : Astrophys. J.           \NC         : Nuovo Cim.
%    \ANN  : Ann. of Phys.           \NPA, \NPB  : Nucl. Phys. [A,B]
%    \CMP  : Commun. Math. Phys.     \PLA, \PLB  : Phys. Lett. [A,B]
%    \IJMP : Int. J. Mod. Phys.      \PRA - \PRE : Phys. Rev. [A-E]     
%    \JHEP : J. High Energy Phys.    \PRL        : Phys. Rev. Lett.
%    \JMP  : J. Math. Phys.          \PRP        : Phys. Rep.
%    \JP   : J. of Phys.             \PTP        : Prog. Theor. Phys.     
%    \JPSJ : J. Phys. Soc. Jpn.      \PTPS       : Prog. Theor. Phys. Suppl.
% Usage:
%  \PRD{45,1990,345}          ==> Phys.~Rev.\ D \textbf{45} (1990), 345
%  \JL{Nature,418,2002,123}   ==> Nature \textbf{418} (2002), 123
%  \andvol{123,1995,1020}    ==> \textbf{123} (1995), 1020
%%%%%%%%%%%%%%%%%%%%%%%%%%%%%%%%%%%%%%%%%%%%%%%%%%%%%%%%%%%%%

\end{document}